\documentclass[prl,priprint,epsf,psfig]{revtex4}
\usepackage{graphicx}
\DeclareGraphicsExtensions{.pdf,.png,.gif,.jpg}
\usepackage{float}
\usepackage{subfig}

\begin{document}

\title{Phase diagram of two-component dipolar fermions in one-dimensional optical lattices}

\author{Theja N. De Silva\footnote{Corresponding author. Tel.: +1 607 777 3853, Fax.: +1 607 777 2546\\
E-mail address: tdesilva@binghamton.edu}}
\affiliation{Department of Physics, Applied Physics and Astronomy,
The State University of New York at Binghamton, Binghamton, New York
13902, USA.}

\begin{abstract}
We theoretically map out the ground state phase diagram of interacting dipolar fermions in one-dimensional lattice. Using a bosonization theory in the weak coupling limit at half filing, we show that one can construct a rich phase diagram by changing the angle between the lattice orientation and the polarization direction of the dipoles. In the strong coupling limit, at a general filing factor, we employ a variational approach and find that the emergence of a Wigner crystal phases. The structure factor provides clear signatures of the particle ordering in the Wigner crystal phases.
\end{abstract}

\maketitle

\section{I. Introduction}

The recent experimental progress in creating degenerate cold polar atoms/molecules with large dipolar moments attracted considerable attention due to the rich quantum mechanical phenomena they can exhibit \cite{edp1, edp2, edp3, edp4, edp5, edp6, edp7, edp8}. In contrast to the contact interaction, the long-range, anisotropic dipolar-dipolar interaction between dipolar molecules offers promising directions for exploring novel and strongly correlated many-body physics. The experimental exploration of dipolar physics started with the observation of Bose-Einstein condensation of $^{52}$Cr and $^{164}$Dy magnetic dipolar atoms \cite{edp1, edp7}. Later, showing promising indication of creating quantum degenerate mixtures of dipolar molecules, a dense gas of $^{40}$K$^{87}$Rb and dual-species Bose-Einstein condensate of $^{87}$Rb and $^{133}$Cs have been realized experimentally \cite{edp2, edp8}. The realization of dipolar molecules in an optical lattice~\cite{add} and the first creation of quantum degenerate dipolar Fermi gas of $^{161}$Dy have just been reported~\cite{dy}. For molecules with permanent electric or magnetic dipole moments, the range of the dipole-dipole interactions can be much larger than typical optical lattice spacings. Optical lattices provide rich tunable ingredients such as geometry, dimensionality, and interactions so that one can engineer novel many-body states~\cite{tut}. These states include various superfluid states such as $p_x +i p_y$ and $d$-wave superfluid phases, supersolid phases, vortex lattices, various Wigner crystal phases, charge-density wave and spin-density wave phases \cite{tdp1, tdp2, tdp3, tdp4, tdp5, tdp6, tdp7, tdp8, tdp9, tdp10, tdp11, tdp12, tdp13, tdp14, tdp15, tdp16, tdp17, tdp18, tdp19, tdp20, tdp21, n1, n2, n3, n4, n5, n6, n7, n8, n9, n10}. Further, cold polar molecules in optical lattices provide a platform for novel spin models and possible applications in quantum computing \cite{sm, qc}.

The physics of non-polar atoms with only contact interaction in optical lattices can be reasonably described by the Hubbard model \cite{jak}. In the Hubbard model, atom-atom interaction is approximated by an on-site interaction $U$. However, as the dipole-dipole interaction is long-range, the experiments of polar atoms or molecules fall outside the range of validity of the Hubbard model. A natural extension of the
Hubbard model comes from including long-range, off-site interactions between the molecules.

The one-dimensional many-body phenomena, such as the break down of Fermi liquid theory and spin-charge separation, can be understood in the framework of bosonization theory \cite{book1, book2, rev3}. The bosonization theory is valid asymptotically at small momenta and low energies. In this Letter, we study the phase diagram of two-component, one-dimensional lattice fermions. While we use the bosonization theory in the weak coupling limit, a variational approach is employed in the strong coupling limit to study the possible Wigner crystal states. Using the bosonization theory at half-filling, we show that one can achieve a rich phase diagram by changing the polarization direction with respect to the lattice orientation. The weak coupling phase diagram includes spin-density wave, charge-density wave, singlet superfluid and triplet superfluid phases. In the strong coupling limit, at smaller filling factors, we find that the long-range interaction induces a Wigner crystal phase. The structure factor or the density-density correlation function which can be measured using Bragg scattering experiments provide a clear signatures of the Wigner crystal phase.

The Letter is organized as follows. In section II, we discuss the effective lattice model for the dipolar fermions in one dimension. In section III, we present the bosonization theory for weakly interacting fermions in the presence of long-range off-site interaction. Assuming that the dipoles are polarized along the applied field and taking the angle between the lattice direction and the applied field as a free parameter, the weak coupling limit ground state phase diagram at half-filling is presented in section IV. The section V is devoted to discuss the effect of inter-chain coupling in realistic experimental settings. In section VI, we consider the strong coupling limit and use a variational approach to study the possible Wigner crystal state away from half-filling. Finally in section VII, a summary is provided.

\section{II. The model}

We consider a system of two-component electric or magnetic dipoles confined in a one-dimensional optical lattice oriented along the $x$-direction. The Hamiltonian operator for the fermionic atoms in optical lattice is given by

\begin{eqnarray}
H = \sum_\sigma \int dx \psi^\dagger_\sigma (x)\biggr[-\frac{\hbar^2}{2m}\frac{\partial^2}{\partial x^2} + V_0(x)\biggr]\psi_\sigma(x)
+\frac{1}{2}V_{ci}\int dx \psi^\dagger_\uparrow (x)\psi^\dagger_\downarrow (x)\psi_\downarrow(x)\psi_\uparrow(x) \\ \nonumber
+ \frac{1}{2}\sum_{\alpha,\beta,\gamma,\delta}\int dx dx^\prime \psi^\dagger_\alpha (x)\psi^\dagger_\beta(x^\prime)\tilde{V}_{dd}(x-x^\prime)\psi_\gamma(x^\prime)\psi_\delta(x),\label{model}
\end{eqnarray}

\noindent where $\psi^\dagger_\sigma(x) [\psi_\sigma(x)]$ is a fermion field operator which creates (annihilates) a Fermi atom with mass $m$ and pseudo-spin $\sigma = \uparrow, \downarrow$ at position $x$. Here the pseudo-spin $\sigma$ refers to the two hyperfine states of the atom. The optical lattice potential provided by the counter propagating laser is $V_0(x) = V_0\sin^2(kx)$, with the wave amplitude $V_0$ and wavevector $k = 2\pi/\lambda$, where $\lambda$ is the laser wavelength corresponding to a lattice period $d = \lambda/2$. The s-wave contact interaction $V_{ci} = 4\pi\hbar^2a_s/m$, with s-wave scattering length $a_s$ and the effective one-dimensional dipolar-dipolar interaction $\tilde{V}_{dd}(x)$, is related to the three-dimensional dipolar-dipolar interaction,

\begin{eqnarray}
V_{dd}(r) = D^2 \frac{1-3 \cos^2 \theta_d}{r^3}\label{dd3D}
\end{eqnarray}

\noindent where $\theta_d$ is the angle between the 1D lattice in the $x$-direction and the dipolar moment of the atoms align along the applied homogeneous electric or magnetic field in the $x-z$ plane. The strength of the dipolar-dipolar interaction is $D^2 = d_0^2/(4 \pi \epsilon_0)$ and $D^2 = \mu_0 d_0^2/(4 \pi)$ for electric and magnetic dipoles respectively. Here $\epsilon_0$ is the electric permittivity, $\mu_0$ is the magnetic permeability, and $d_0$ is the dipolar moment. For a tight one-dimensional geometry, the level spacing in transverse direction is much larger than the energy per particle of the axial direction $x$. The integration of the dipolar-dipolar interaction in Eq. (\ref{dd3D}) over the transverse direction leads to the effective one-dimensional dipolar interaction \cite{tdp6, tdp9}

\begin{eqnarray}
\tilde{V}_{dd}(x) = -D^2 \frac{1+3 \cos (2\theta_d)}{x^3}.\label{dd1D}
\end{eqnarray}

The single atomic energy eigenstates are Bloch states and localized Wannier functions are a superposition of Bloch states. For a deep optical lattice with atoms trapped in the lowest vibrational states $w(x) = e^{-x^2/(2l^2)}/\sqrt{l\pi^{1/2}}$ with $l = \sqrt{\hbar/(m\omega)}$, the field operators $\psi_{\sigma}$ can be expanded as $\psi_\sigma = \sum_ic_{i\sigma} w(x-x_i)$. The oscillator length $l$ is defined through $\hbar \omega = \sqrt{4E_RV_0}$. Here $\omega$ is the oscillation frequency, obtained using the harmonic approximation around the minima of the optical potential well at each lattice site. The recoil energy is $E_R = \hbar^2k^2/(2m)$. In terms of new fermionic operators $c_{i\sigma}$ at lattice site $i$, the effective lattice Hamiltonian for the polar fermionic system reads \cite{tut, tdp14},

\begin{eqnarray}
H = -t \sum_{\langle ij \rangle,\sigma} (c_{i\sigma}^\dagger c_{j\sigma} + h. c) + U \sum_i n_{i\uparrow}n_{i\downarrow} + \sum_{i,r} V_{ir} n_{i+r}n_{i}.\label{model2}
\end{eqnarray}

\noindent The parameters $t = \int dx w^\ast(x-x_i)[-\frac{\hbar^2}{2m}\frac{\partial^2}{\partial x^2}+ V_0(x)]w(x-x_j)$ is the hopping matrix element between neighboring sites $i$ and $j$, $U = 4\pi\hbar^2a_s \int dx |w(x)|^4/m$ is the on-site interaction of the two atoms at site $i$, and $V_{ir} = \int dx dx^\prime |w(x-x_i)|^2 \tilde{V}_{dd}(x-x^\prime) |w(x^\prime-x_r)|^2$ is the off-site interaction of two atoms at sites $i$ and $r$. Apart from this "direct" like off-site density-density interaction term, "exchange" like spin-spin interaction term is also present for dipolar gases~\cite{kaden}. Assuming, dc electric and microwave fields in the realistic experimental setups allow one to tune the "direct" like interactions to be dominant~\cite{alex}, here we neglect the spin-spin interaction term. In terms of $\omega$, $V_0$, and $l$, the parameters read, $t =  e^{-\pi^2 V_0/(2\hbar\omega)}\hbar\omega/2$, $U = 4\pi\hbar^2a_s/(\sqrt{2\pi} ml)$, and $V_{ir} = -V [1+3 \cos(2 \theta_d)]/(|i-r|^3)$. Notice that the tunneling energy is exponentially sensitive to the laser intensity whereas the interactions are weakly sensitive. The on-site interaction can be either repulsive or attractive depending on the sign of the scattering length $a_s$. Furthermore, off-site interaction $V_{ir}$, can be adjusted to be positive (repulsive) or negative (attractive) by changing the direction of the applied field. Here $i-r \neq 0$ is a discrete variable that represents the lattice points. We consider both repulsive and attractive regimes under the assumption that purely attractive regime is achievable in the metastable state as has been experimentally demonstrated for one-dimensional bosonic Cs atoms~\cite{cs}. Perhaps, the residue of small spin-spin interaction term restore the mechanical stability in the attractive regime.

\section{III. Bosonization theory}

For asymptotic low-energy properties and for the weak coupling regime of the system, the continuum limit is a good approximation. We use the standard bosonization techniques \cite{bosT} to map the Hamiltonian into the continuum limit by introducing continuous fermion fields $c_{i\sigma}/\sqrt{d} \rightarrow \psi_{L\sigma}(x) + \psi_{R\sigma}(x)$ with

\begin{eqnarray}
\psi_{\eta\sigma}(x) = \frac{U_{\eta\sigma}}{\sqrt{2\pi\alpha}} e^{i\eta k_F x}e^{i/\sqrt{2} [\eta (\phi_n + \sigma \phi_\sigma)-\theta_n-\sigma \theta_\sigma]}.\label{e5}
\end{eqnarray}

\noindent The fermion operator $\psi_{\eta\sigma}^\dagger(x)$ creates a Fermi atom of pseudo-spin $\sigma$ on the branch $\eta = R, L = \pm 1$ of the linearized spectrum $E(k) = v_F(\eta k-k_F)$, where $v_F = 2dt\sin(k_Fd)$ is the Fermi velocity. Here $R$ and $L$ refer to right movers and left movers, respectively. The parameter $\alpha$ is the standard bosonization short-range distance cut-off, which is on the order of a lattice constant $d$. The Fermi wavevector is $k_F = \pi n/(2d)$ with particle density $n$. In the continuum limit, $x = jd$ and the length of the chain $L = Nd$ is finite, hence we consider the limits $d \rightarrow 0$ and the number of atoms $N \rightarrow \infty$. The discrete variable $j$ above represents the lattice points. The fields representing particle($\nu =n$) and spin($\nu = \sigma$) fluctuations are $\phi_\nu$ and $\theta_\nu$. They satisfy the commutator $[\phi_\mu(x), \theta_\nu(x^\prime) ] = -i\pi/2 \delta_{\mu,\nu}sgn(x-x^\prime)$. The Hermitian operators $U_{\eta\sigma}$ satisfy the commutator $[U_{\eta\sigma}, U_{\eta^\prime\sigma^\prime} ] = 2 \delta_{\eta\eta^\prime}\delta_{\sigma \sigma^\prime}$. Introducing the velocities $v_\nu$ of particle ($n$) and spin ($\sigma$) sectors and Gaussian couplings $K_\nu$, and following the standard procedure, the 1D particle system can be represented by the sine-Gordon model as \cite{sam1, joha},

\begin{eqnarray}
H = \sum_{\nu =n, \sigma} \frac{v_\nu}{2\pi}\int_0^L dx \biggr[K_\nu (\partial_x\theta_\nu)^2 + \frac{1}{K_\nu}(\partial_x\phi_\nu)^2\biggr] \\ \nonumber + \frac{2g_{1\perp}}{(2\pi\alpha)^2}\int_0^L dx \cos[\sqrt{8} \phi_\sigma(x)] \\ \nonumber
+ \frac{2g_{3\perp}}{(2\pi\alpha)^2}\int_0^L dx \cos[q\sqrt{8} \phi_n(x) + \delta x] \\ \nonumber
+ \frac{2g_{3\parallel}}{(2\pi\alpha)^2}\int_0^L dx \cos[q\sqrt{8} \phi_n(x)+ \delta x]\cos[q\sqrt{8} \phi_\sigma(x)].\label{BH}
\end{eqnarray}

\noindent Here we use the standard notations where

\begin{eqnarray}
v_\nu &=& v_F[(1+y_{4\nu}/2)^2-(y_\nu/2)^2]^{1/2} \\ \nonumber
K_\nu &=& \biggr[\frac{1+y_{4\nu}/2+y_\nu/2}{1+y_{4\nu}/2-y_\nu/2}\biggr]^{1/2} \\ \nonumber
g_\nu &=& g_{1\parallel}-g_{2\parallel} \mp g_{2\perp} \\ \nonumber
g_{2\nu} &=& g_{2\parallel}\pm g_{2\perp} \\ \nonumber
g_{4\nu} &=& g_{4\parallel} \pm g_{4\perp} \\ \nonumber
y_\nu &=& g_\nu/(\pi v_F) ,\label{e7}
\end{eqnarray}

\noindent where the upper sign refers to the particle sector ($n$) and the lower sign refers to the spin sector ($\sigma$). In standard bosonization language, the coupling constants $g_{i\parallel}$ and $g_{i\perp}$ with $i =1,...4$, refer to low-energy processes of the interaction. The coupling $g_1$ couples two fermions on the opposite side of the Fermi surface and the particles switch the sides after the interactions. This process is called backward scattering or $2k_F$ scattering. The coupling $g_2$ couples two fermions on the opposite sides of the Fermi surface which stay on the same side after the scattering. This process is called forward scattering. Notice that the effect of $g_2$ is included in the first term in Eq. (6)~\cite{joha}. The coupling between two fermions on the same side of the Fermi surface is denoted by the coupling constant $g_4$. The subscripts $\parallel$ and $\perp$ refer scattering between fermions with parallel spins and anti-parallel spins, respectively. The scattering corresponding to the coupling constants $g_{3\perp}$ and $g_{3\parallel}$ occurs only in the presence of the lattice. These are the well-known umklapp processes where the momentum is conserved up to the reciprocal lattice vector. The parameters $\delta$ and $q$ control the filling factor $n = N/L$. In this section of the present Letter we treat the half-filling case where $\delta =0$ and $q=1$ so that $n = 1$.

In the weak coupling limit of our model in Eq. (6), all the scattering amplitudes can be presented as follows \cite{sam1, sam2, sam3}. The amplitudes of the backward scattering are $g_{1\perp} = Ud +2d\sum_xV_x \cos(2k_Fx)$ and $g_{1\parallel} = 2d\sum_xV_x \cos(2k_Fx)$. Notice that we use $V_{ir} \rightarrow V_x$ to represent the discrete variable $|i - r| \rightarrow dx$. The amplitudes of the forward scattering are $g_{2\perp} = Ud - 2d\sum_xV_x \cos(2k_Fx)$ and $g_{2\parallel} =-g_{1\parallel}$. The amplitudes of the umklapp scattering are $g_{3\perp} = g_{1\perp}$ and $g_{3\parallel} = g_{1\parallel}$. The amplitudes of the other scattering are $g_{4\perp}= g_{2\perp}$ and $g_{4\parallel}= g_{2\parallel}$. For the case of weak coupling, the velocities and the Gaussian coupling in the particle and spin sectors are $v_\nu K_\nu = v_F$, $v_n/K_n = v_F - g_n/\pi$, and $v_\sigma/K_\sigma = v_F - g_\sigma/\pi$.

\section{IV. Phase diagram}

In the absence of the umklapp processes and in the limit $g_{1\perp}\rightarrow 0$, the Hamiltonian is quadratic. In this limit, various correlation functions corresponding to different quantum phases can be easily calculated. These correlation functions show non-universal power law decay with exponents depending on Gaussian couplings $K_n$ and $K_\sigma$ \cite{joha}. However, in the presence of umklapp processes and the non-zero limit of $g_{1\perp}$, one has to treat the quantum phase transitions by using renormalized group theoretical techniques.

\begin{figure}
\includegraphics[width=\columnwidth]{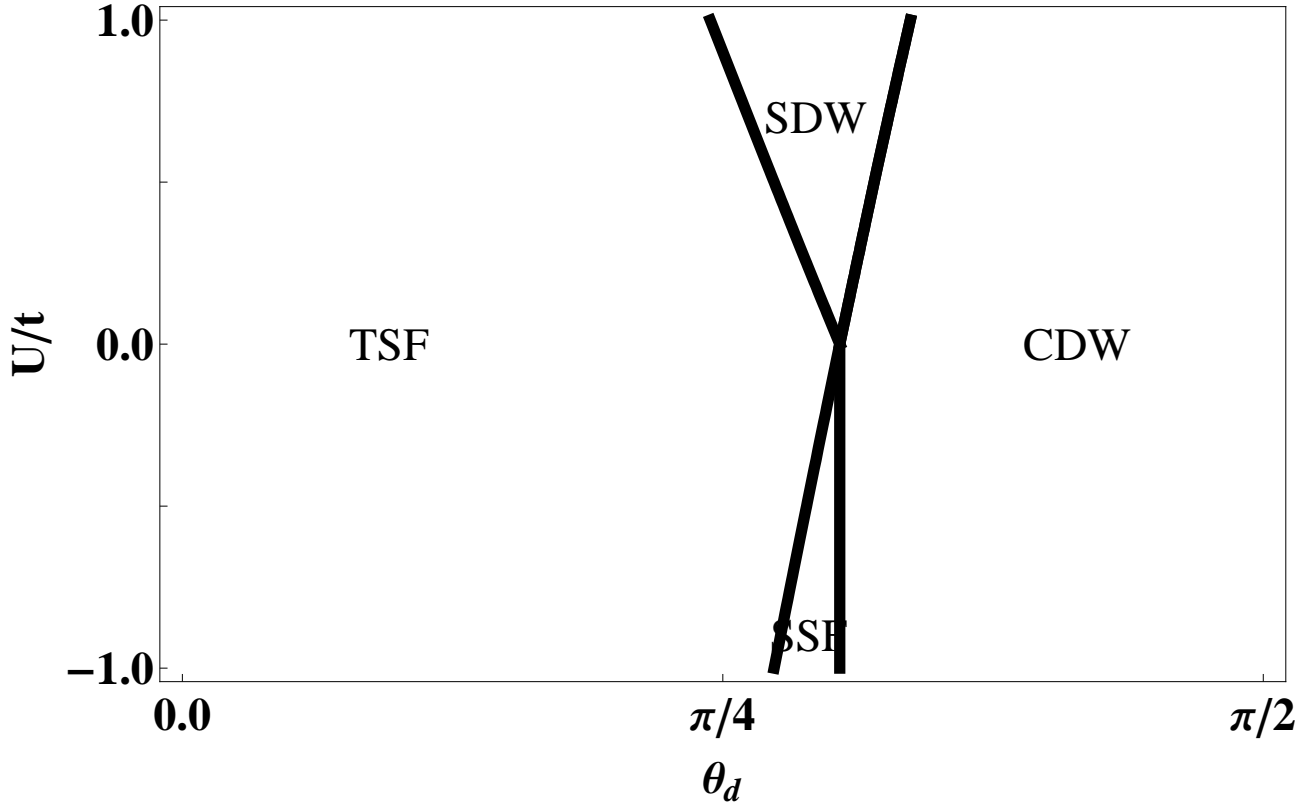}
\caption{The phase diagram of one-dimensional polarized dipolar fermions in the weak coupling limit. The angle $\theta_d$ is the polarized angle with respect to the 1D lattice orientation. We set the filling factor to $n = 1$ and the long-ranged dipolar interaction up to 100 lattice sites. The abbreviated phases are SDW: spin-density wave, CDW: charge-density wave, TSF: triplet superfluid phases and  SSF: singlet superfluid.} \label{pd}
\end{figure}

\begin{figure}
\includegraphics[width=\columnwidth]{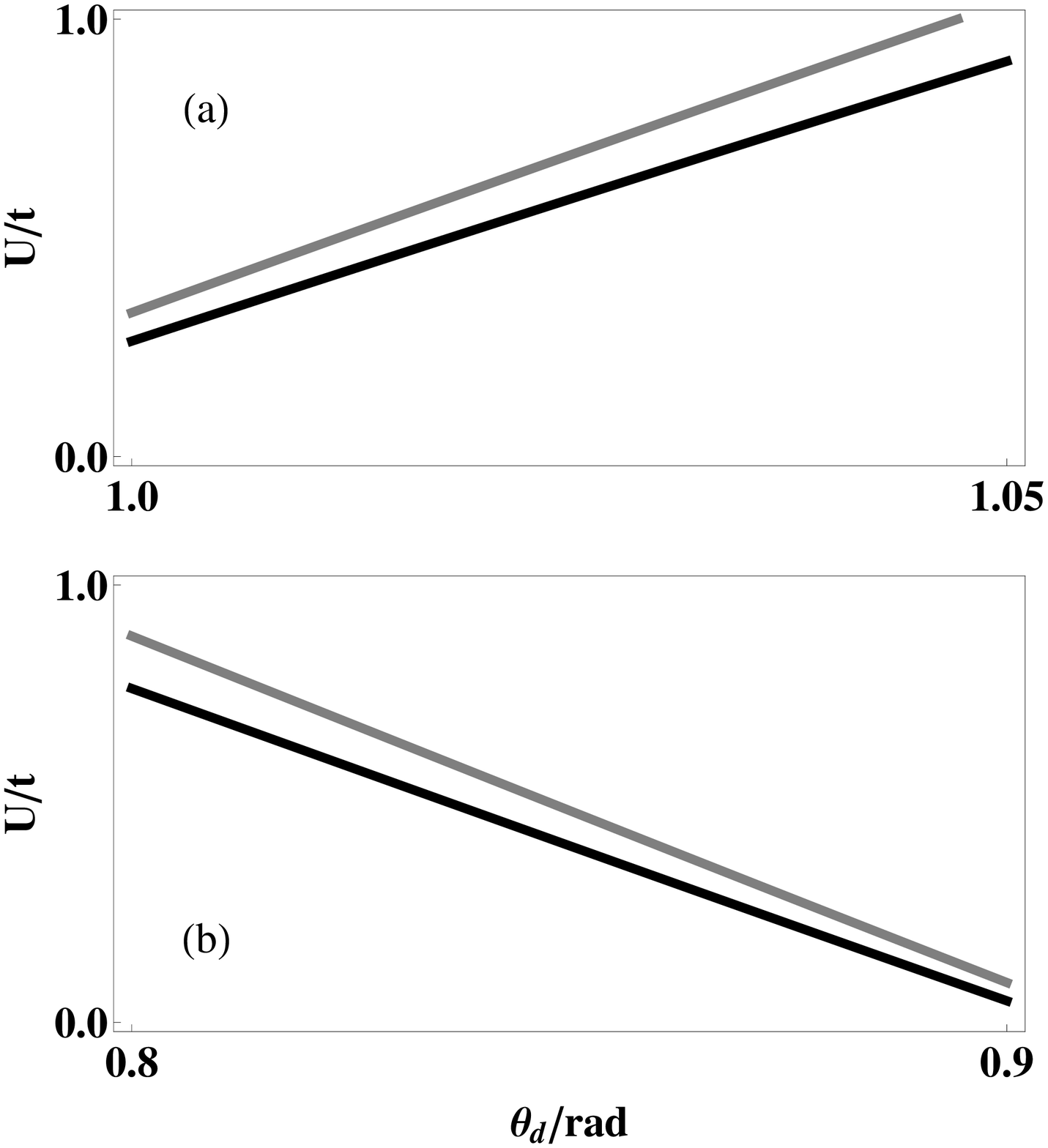}
\caption{The effect of the interaction range. Panel (a) shows the boundary line between a charge-density wave phase and a spin-density wave phase for $m = 50$ (black) and $m = 1$ (gray). Panel (b) shows the boundary line between a spin-density wave phase and a triplet superfluid phase for $m = 50$ (black) and $m = 1$ (gray). Notice that $m =1$ represents only the nearest neighbor interaction.} \label{range}
\end{figure}

In the present section, we consider weak coupling limit at half-filling. In the weak coupling limit, the scaling dimension of $g_{3\parallel}$ term is always higher than that of other non-linear terms in our model~\cite{sam1, joha}. Therefore, we set $g_{3\parallel} =0$ in the present study. The effect of $g_{1\perp}$ and $g_{3\perp}$ is taken into account using renormalized group (RG) equations as has been done in Ref. \cite{sam1}. Changing the cut-off $\alpha \rightarrow e^{dl} \alpha$ with $l = \ln L$, the RG equations within one-loop order is given by

\begin{eqnarray}
\frac{dy_{\nu0}(l)}{dl} &=& -y^2_{\nu\phi}(l) \\ \nonumber
\frac{dy_{\nu \phi}(l)}{dl} &=& -y_{\nu 0}(l) y_{\nu \phi}(l)\label{RGF}
\end{eqnarray}

\noindent where $y_{n 0}(0) = 2 (K_n -1)$, $y_{\sigma 0}(0) = 2 (K_\sigma -1)$, $y_{n \phi}(0) = g_{3\perp}/(\pi v_n)$, and $y_{\sigma \phi}(0) = g_{1\perp}/(\pi v_\sigma)$. Notice that we consider the weak coupling limit at half-filling. At these limits, the RG equations for particle and spin sectors are decoupled~\cite{joha}. These equations determine the RG flow diagrams as presented in FIG. 2 of ref. \cite{sam1}. The RG equations for velocities at these limits are trivial and velocities have no effect on scaling dimension. Following the same arguments as in ref. \cite{sam3}, the weak coupling phase diagram at half-filling can be extracted from the RG flow diagram.

For the spin-gap transition ($\nu =\sigma$), Eq. (8) gives

\begin{eqnarray}
y_{\sigma 0}(l) = \frac{ly_{\sigma 0}(0)}{l y_{\sigma 0}(0)+1}.\label{e9}
\end{eqnarray}

\noindent This shows that the spin gap opens when $y_{\sigma 0}(l) < 0$. For the weak coupling limit, where $g_\nu/(\pi v_F) \ll 1$, Gaussian coupling $K_\nu = [1-g_\nu/(\pi v_\nu)]^{-1/2}$ can be approximated as $K_\nu \simeq 1+ g_\nu/(2 \pi v_\nu)$. In this limit, the condition $y_{\sigma 0}(l) < 0$ translates into $g_\sigma < 0$. Therefore, the phase boundary between the charge-density wave (CDW) and the spin-density wave (SDW) phases at half-filling is determined by $U = 4V[1 + 3\cos (2 \theta_d)] \sum_{m =1}(-1)^m/m^3$. The sum over $m$ controls the range of the long-range interaction. For example, $m =1$ represents only the nearest neighbor interaction.

On the other hand, the condition for the charge gap is $g_{3\perp} > |g_n|$. This condition gives two possible phase boundaries; one is $g_{3\perp} = -g_n < 0$ and the other is $g_{3\perp} = g_n > 0$. As there is no continuous symmetry breaking in one dimension, these phase transitions, due to the opening of a charge gap, are not true phase transitions but have power law decaying correlations. These Berenzinskii-Kosterlitz-Thouless type transitions are due to the SU(2) and hidden SU(2) symmetries in the particle (charge) sector \cite{bkt1, bkt2}.

The boundary between CDW phase and the singlet super-conducting correlation (SSF) phase is given by the conditions $U < 0 $ and $\sum_x V[1+3 \cos (2\theta_d)]\cos(2k_Fx) =0$. At half-filling, this condition translates into $U < 0$ and $\cos (2\theta_d) = -1/3$.

The phase boundary between SDW phase and the triplet supper-conducting correlation (TSF) phase is determined by the conditions $U > 0$ and $U = -4V [1+3 \cos(2\theta_d)]\sum_{m=1} (-1)^m/m^3$. Similar to the quadratic Hamiltonian \cite{joha}, a Gaussian type transition take place between two gapped phases when $y_{n\phi} = 0$ and $y_{n0} <0$. Since non-linear term vanishes on this Gaussian type transition line, this transition between SSF and TSF phase does not emerge from the RG equations~\cite{sam1}. Instead, the scaling dimensions on the Gaussian line determines the transition at $g_n < 0$ at $g_{3\perp} = 0$. Therefore, the phase boundary between SSF and TSF phases are given when $U = 4V [1+3 \cos(2\theta_d)] \sum_{m =1}(-1)^m/m^3$ and $\cos(2\theta_d) < -1/3$. The resulting phase diagram in $U-\theta_d$ plane for $2V =1 $ is shown in FIG. \ref{pd}.

All the phases in the rich phase diagram in FIG. \ref{pd} can be constructed experimentally just by fixing the on-site interaction (i.e fixing the laser intensity of the counter propagating lasers) and then changing the polarization direction with respect to the lattice orientation. However, the interactions have to be weak and the filling factor must be unity. By controlling the total number of particles in experiments, the filling factor at the center of the lattice can be set to unity. In-situ density imaging or Bragg spectroscopy can be used to distinguish the charge-density wave and the spin-density wave phases. The superfluid phases can be detected via pair correlation measurements using noise spectroscopy \cite{alt}.

Notice that the boundary between the singlet superfluid and charge-density wave phases does not depend on the range of the interaction. However, all of the other boundaries have an effect on the range of long-range dipolar interaction. The shift of the boundaries due to the long-range part of the interaction is shown in FIG. \ref{range} for a fixed $V = 0.5$. The interaction strength $V$ is always positive so that the qualitative features of the phase diagram do not change with $V$.

\section{V. The effect of inter-chain coupling}

In the field of cold-atomic physics, one-dimensional systems are realized by creating an array of many 1D-tubes. Even though the tunneling between tubes is absent for well separated chains, the long-range dipolar-dipolar interaction can still cause coupling between tubes. In the presence of inter-chain coupling, we must modify our original Hamiltonian in Eq. (1) by adding the inter-chain interaction term,

\begin{eqnarray}
 H_I = \frac{1}{2}\sum_{\alpha,\beta,\gamma,\delta,j, j^\prime}\int dx dx^\prime \psi^\dagger_{j,\alpha} (x)\psi^\dagger_{j^\prime,\beta}(x^\prime)V_{dd}(x-x^\prime)\psi_{j,\gamma}(x^\prime)\psi_{j^\prime\delta}(x),\label{e10}
\end{eqnarray}

\noindent where $j \neq j^\prime$ is the chain index. Generalizing the 1D Wannier function $w(x) \rightarrow w(x, y) = e^{-[x^2 + (y-j R)^2]/(2l^2)}/\sqrt{l\pi^{1/2}}$, the inter-chain coupling can be approximated by $V_\perp = 2 D^2 \sin\theta^2_d/R^2$, where we assume that the neighboring chain is $R$ distance away in the $y$-direction. In the absence of the lattice, the planar array of 1D tube systems has been studied using bosonization theory \cite{tubes}. It has been shown that the inter-chain interaction is irrelevant, except when $\theta_d \simeq \theta_c$ where the long-range interaction vanishes along the lattice in the $x$-direction. When $\theta_d \simeq \theta_c$, the 1D system approaches the boundary between CDW and SSF phases and the long-range positive interaction between neighboring chains induces a type of CDW phase in the transverse direction \cite{tubes}. Even in the presence of a lattice, the inter-chain coupling can induce an inter-chain CDW phase. As a result, the intra-chain CDW-SSF phase boundary shifts toward the CDW phase allowing the SSF phase to stabilize over the CDW phase into a larger region of the phase diagram.

\section{VI. The strong coupling limit}

As has been shown above, for any positive on-site and off-site interactions, the particle system produces a insulating phase at half-filling. For any commensurate filling factors away from half-filling, the umklapp scattering is an irrelevant perturbation \cite{umk}. For the 1D Hubbard model without the off-site interactions, the system remains in a metallic phase at any filling factor away from half-filling. However in the presence of long-range interactions, when the average particle spacing $1/n$ is comparable to the range of the interaction, atoms may form a self-organized pattern known as a Wigner crystal phase. The transition into this insulating phase occurs at Luttinger parameter $K_n = n^2$.

\begin{figure}
\includegraphics[width=\columnwidth]{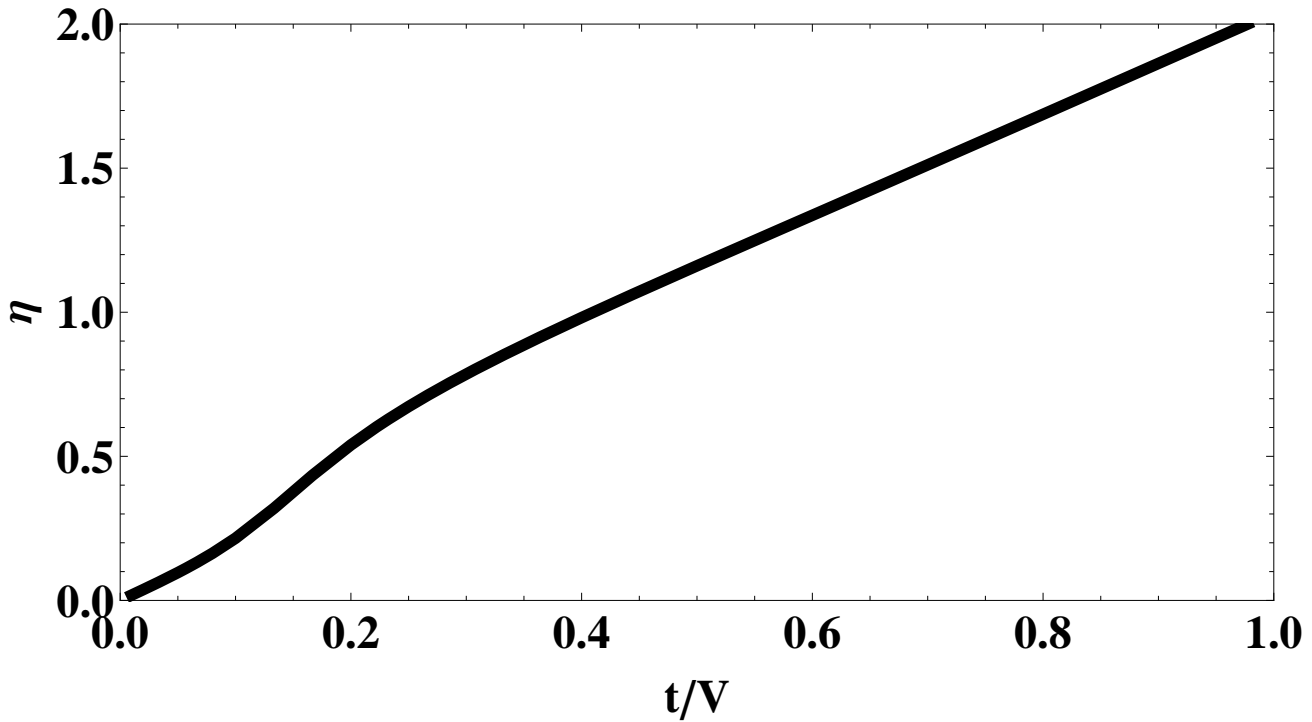}
\caption{The variational parameter $\eta$ for the quarter filling case.} \label{eta}
\end{figure}

\begin{figure*}
\centering
\subfloat[Filling factor $n = 1/2$]{\includegraphics[width=0.5\columnwidth]{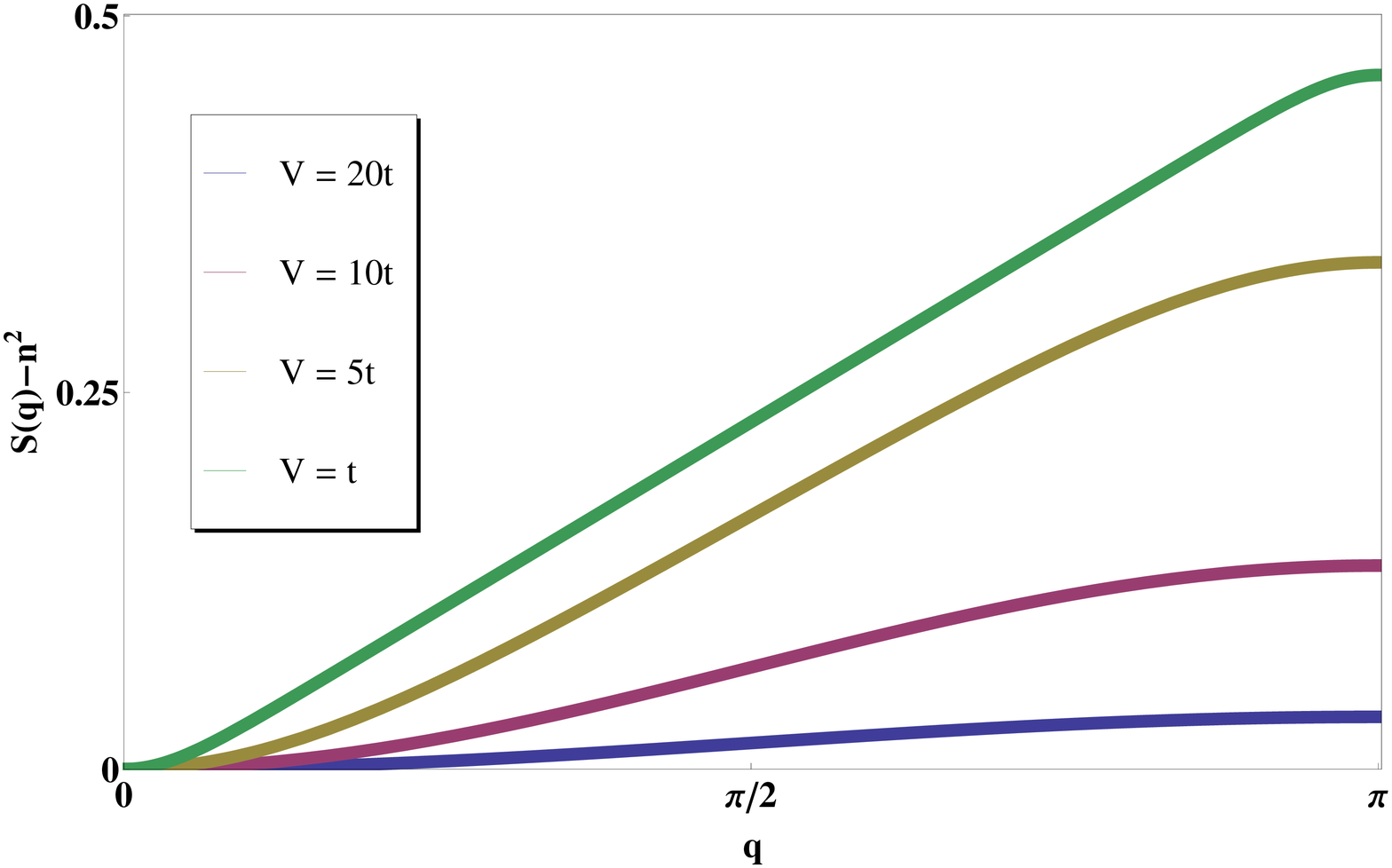}}
\subfloat[Filling factor $n = 1/4$]{\includegraphics[width=0.5\columnwidth]{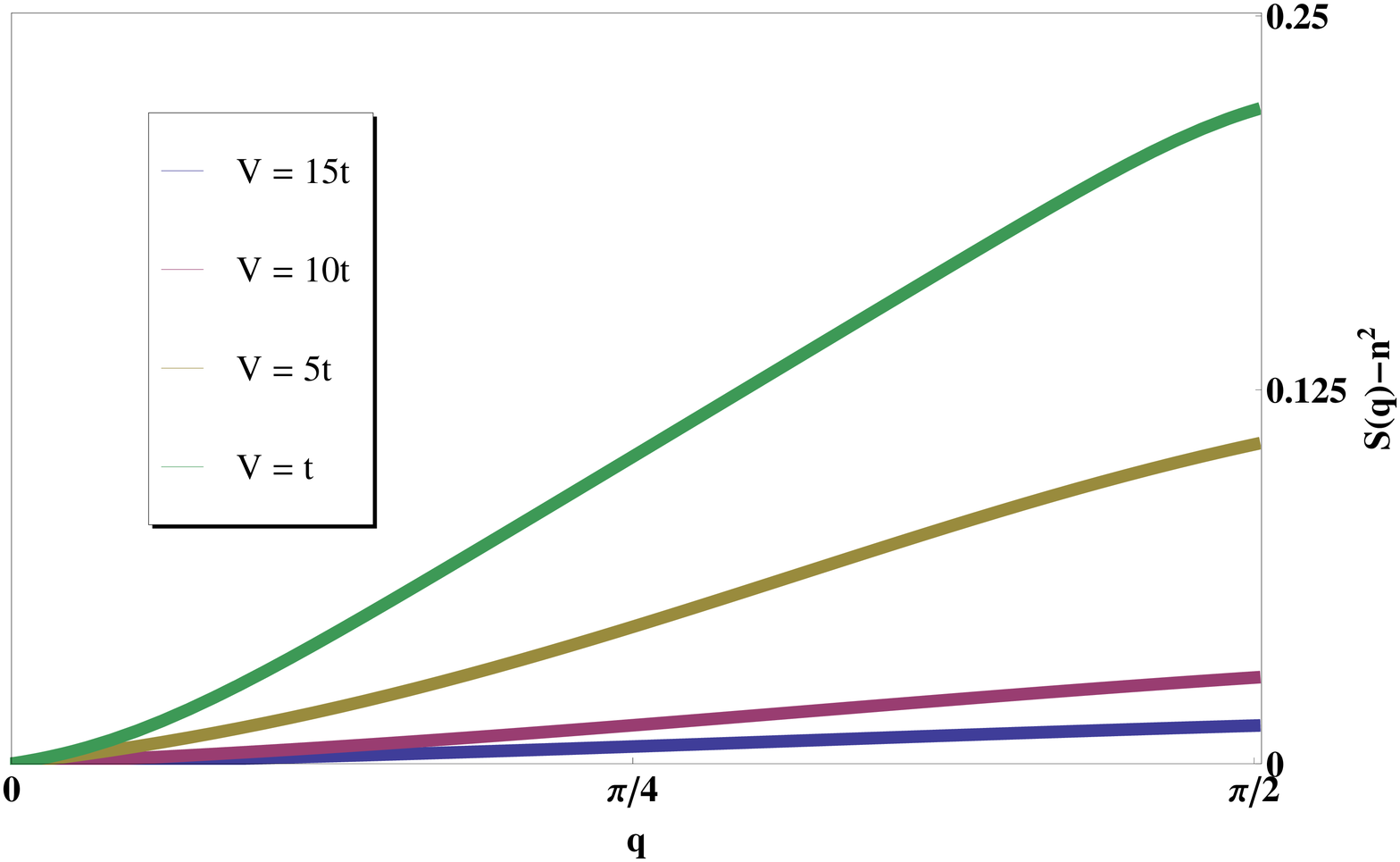}}
\caption{[Color online] The evolution of the structure factor for different interactions for two different filling factors.} \label{Sfactor}
\end{figure*}
For a continuous 1D system, the form of $1/r^\beta$ interaction has been studied in bosonization theory by treating the long-range forward scattering as a perturbation \cite{tsu}. This study shows that for $\beta > 0$, the forward scattering is an irrelevant perturbation. For $\beta = 1$, the Fourier transform of the interaction has a logarithmic divergence. At this limit, the bosonization theory predicts the existence of quasi-Wigner crystal phases with $4k_F$ density correlations \cite{tsu}. At low filling factors, these $4k_F$ correlations are dominant over the $2k_F$ Friedel density correlations. The study in Ref. \cite{tsu} is based on a perturbation approach. Therefore, the existence of quasi-Wigner crystal at larger interactions for $\beta = 3$ is not ruled out. Indeed, by investigating the structure factor using an exact diagonalization method, the existence of Wigner crystals at strong coupling limits and lower filling factors has been verified in Ref. \cite{tdp14}. This verification can be justified by the conditions that the dipolar-dipolar interaction $V_{dd}(r) > 0$, $V_{dd}(r) \rightarrow 0$ as $r \rightarrow \infty$, and $V_{dd}(r+1) + V_{dd}(r-1) \geq 2 V_{dd}(r)$ for any $r >1$. Since the bosonization techniques rely on a linear band dispersion, low energies and long wavelengths, our approach above does not predict the Wigner crystal phase even away from half-filling.

In order to study the possible existence of a Wigner crystal phase at low filling factors, we use a variational approach. Let's set $\theta_d =0$ so that the dipolar-dipolar interaction is purely repulsive. We consider the strong coupling regime where $U$, $V_{ir} \gg t$. As our motivation is to study whether Wigner crystal phases are favorable due to the long-range interaction, we consider the limit $U \rightarrow \infty$. In this limit, both double occupancy and mixing of different spin configurations are eliminated so that we can neglect the local interaction term and suppress the spin index of the operators in Eq. (\ref{model2}). In other words, the Wigner crystal phase associates with charge ordering at this limit can be described as spinless fermions on the lattice. In Fourier space, the resulting Hamiltonian reads,

\begin{eqnarray}
H = \sum_k \epsilon_k c^\dagger_k c_k + \frac{1}{2L}\sum_q V(q) n_q n_{-q},\label{e11}
\end{eqnarray}

\noindent where $\epsilon_k = -2t\cos(kd)$, $n_q = \sum_k c^\dagger_{k+q}c_k$, and $V(q) = 2(qd) K_1(dq)$ is the Fourier transform of the off-site interaction of the form $V_{i-j} = V [L/\pi \sin(\pi |i-j|/L)]^{-3}$. Here $K_1(x)$ is the modified first order Bessel function \cite{ino}. The open boundary conditions in realistic cold atoms systems may cause edge localization phenomena if the number of lattice sites is small~\cite{obc}. However, in the presence of large number of lattice points, we believe that these effects are absent. Therefore, we assume that the optical lattice obeys periodic boundary conditions and introduced a chord distance between sites $i$ and $j$ to include the periodic boundary conditions. We consider any commensurate filling factors in the form $n = N/L = 1/s$ so that one dipolar particle is occupied in a periodic sequence of unit cells of size $s$ in the Wigner crystal phase. Following Ref. \cite{vale}, we take our variational wave function in the form $|\psi(\eta) \rangle = \exp [-\eta \hat{T}] |\psi_0 \rangle$, where, $\hat{T} = -1/(t) \sum_k \epsilon_k c^\dagger_kc_k$, $\eta$ is the variational parameter and

\begin{eqnarray}
|\psi_0 \rangle = \prod_{k \in RBZ} \frac{1}{N_k}(c^\dagger_k + c^\dagger_{k+Q})|0\rangle.\label{e12}
\end{eqnarray}

\noindent Here $Q = 2\pi/s$, $|0\rangle$ is the vacuum state, and $RBZ$ stands for Reduced Brillouin Zone. This wave function is analogous to the well- known Gutzwiller wave function \cite{gutz} which is used to explain the metal-Mott-insulator transition at half filling. Similar to the suppression of doubly occupied states in Gutzwiller wave function, the exponential operator sitting in front of our variational wave function suppresses high kinetic energy states. The normalization factor $N_k^2 = \exp[-2\eta \epsilon_k/t] + \exp[-2\eta \epsilon_{k+Q}/t] \equiv A_k^2 + B_k^2$. Notice that $|\psi_0 \rangle$ is proportional to the classical ground state of the Wigner crystal phase in the absence of tunneling between sites. The variational parameter $\eta$ is determine by minimizing the ground state energy $E_g = \langle\psi(\eta)|H|\psi(\eta)\rangle \equiv \langle \hat{KE} \rangle + \langle \hat{V} \rangle$, where the first term is the kinetic energy and the second term is the off-site interaction energy. By converting the sums in to integrals over the RBZ, the variational kinetic and interaction energies take the form,

\begin{eqnarray}
\langle \hat{KE} \rangle &=& \int_{RBZ} \frac{dk}{2\pi}\frac{\epsilon_k A_k^2+\epsilon_{k+Q} B_k^2}{N_k^2} \\ \nonumber
\langle \hat{V} \rangle &=& \int_{FBZ} \frac{dq}{4\pi}S(q) V(q). \label{e13}
\end{eqnarray}

\noindent The structure factor $S(q) = \langle n_q n_{-q} \rangle$ above has the form

\begin{widetext}
\begin{eqnarray}
S(q) &=& \biggr(\frac{Q}{2\pi}\biggr)^2 + \biggr(\frac{Q}{2\pi}\biggr) - \int_{RBZ} \frac{dk}{2\pi} \frac{A_k^2 A_{k-q}^2 + B_k^2 B_{k-q}^2}{N_k^2 N_{k-q}^2} - \int_{RBZ} \frac{dk}{\pi} \frac{A_k^2 B_k^2}{N_k^4}.\label{e14}
\end{eqnarray}
\end{widetext}

\noindent Notice that the $q$ sum in the interaction energy run over the entire Brillouin zone (FBZ) including $q =0$. This is different from electronic systems where the $q =0$ term is omitted due to the divergency of interaction \cite{wigner}. The positive background charges in the electronic lattice ensures the cancelation of this divergency. As a demonstration, the variational parameter $\eta$ for the quarter filling case ($n = 1/2$) is shown for different values of the interaction strengths in FIG. \ref{eta}. Notice that the variational parameter reaches the classical Wigner crystal phase limit ($\eta = 0$) for larger interaction strengths while it reaches the liquid phase value ($\eta \rightarrow \infty$) in the opposite limit. As our variational wave function always represents the Wigner crystal phase, our approach does not allow us to study the phase transition between Wigner crystal and liquid phases. The qualitative behavior of $\eta$ is similar for other filling factors, however as the filling factor decreases, the variational parameter $\eta$ increases. A justification of phase transition from a liquid phase to a Wigner crystal phase is provided at the end of this section.

\begin{figure}
\includegraphics[width=\columnwidth]{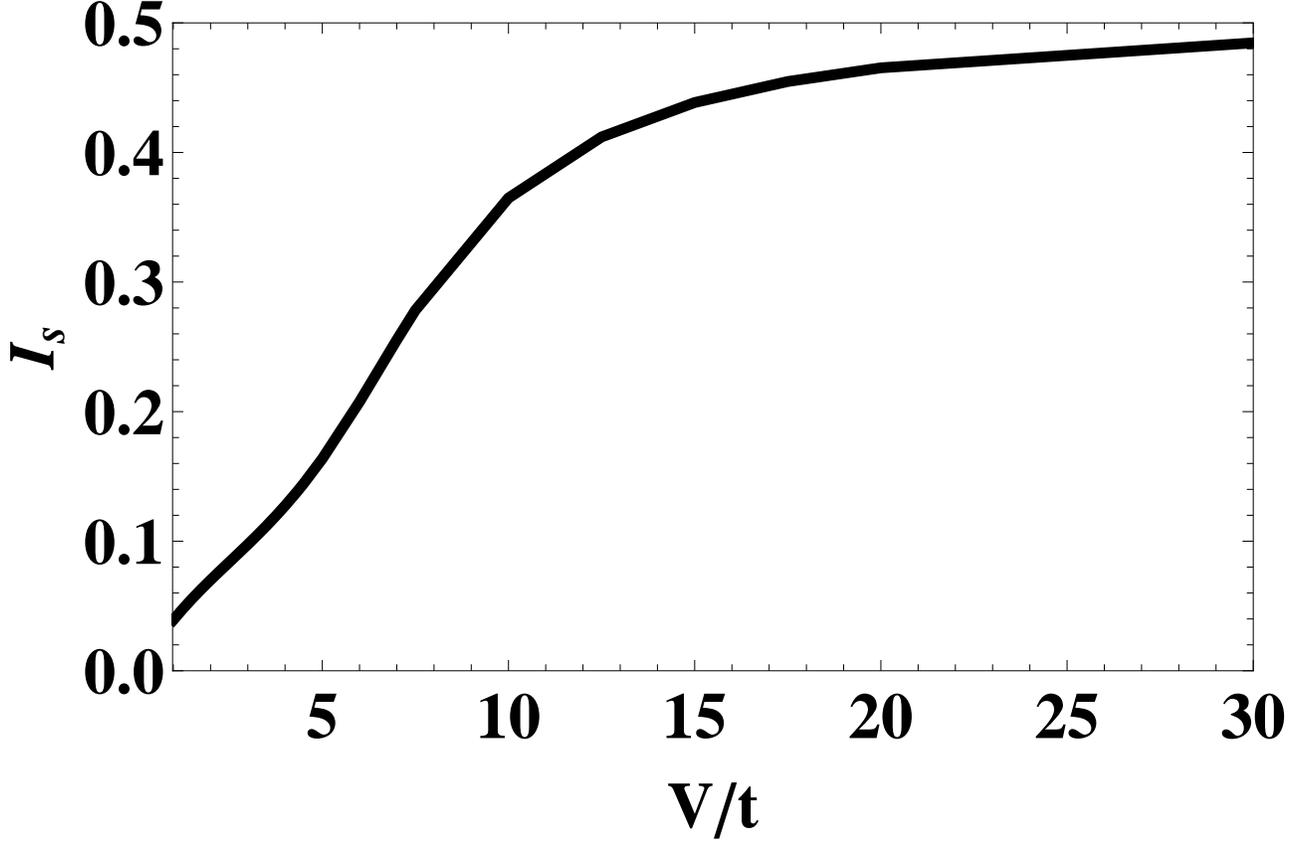}
\caption{The weight of the Bragg peak ($I_s$) for the quarter filling case.} \label{Bpeak}
\end{figure}

As the density-density correlation function is related to the structure factor, the evolution of the structure factor in FIG. (\ref{Sfactor}) shows how the particle modulation builds up as one increases the interaction. We have shown the results for two filling factors $n =1/2$ and $n = 1/4$ corresponding to $Q = \pi$ and $Q = \pi/2$, respectively. The qualitative behavior for other filling factors are the same. The reduction of the structure factor at higher interaction is due to the transfer of some of its weight to the Bragg peak at $q = Q$. The weight transferred to the Bragg peak ($I_s$) can be calculated using $I_s = n-[S(Q)-n^2]$, where $n = Q/(2\pi)$. This peak intensity as a function of the interaction for the quarter filling case is shown in FIG. \ref{Bpeak}. As one expects, the peak intensity goes to zero for non-interacting systems. Since our variational approach is valid only for the Wigner crystal phase, the peak intensity is always non-zero for any finite interactions. However, if the Wigner crystal phase is absent, then the peak intensity must be zero. Experimentally, these Bragg peaks can be probed by measuring the structure factor using Bragg scattering or imaging techniques \cite{bs1, bs2, bs3, bs4, bs5, bs6}.

\begin{figure*}
\centering
\subfloat[Filling factor $n = 1/2$]{\includegraphics[width=0.5\columnwidth]{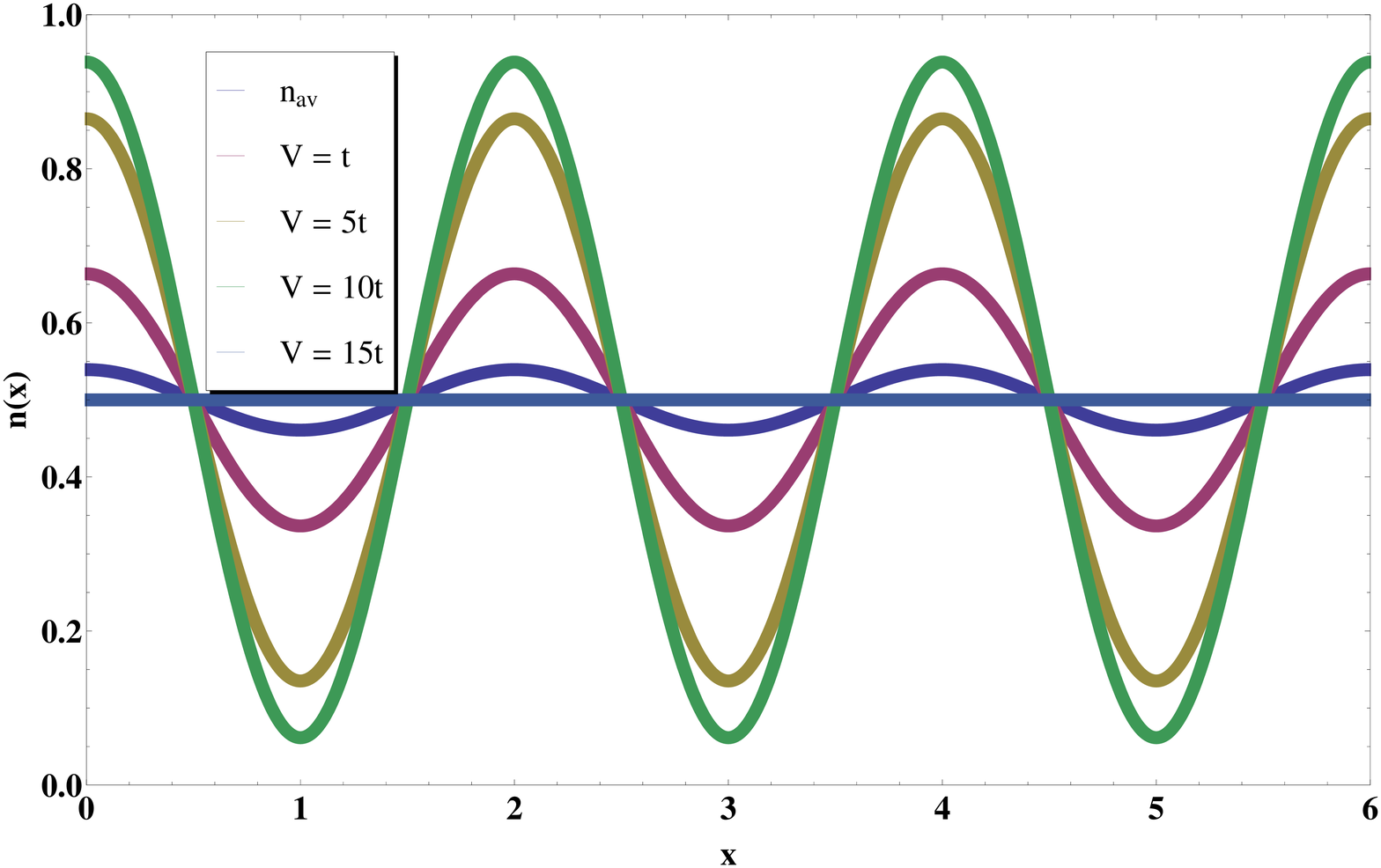}}
\subfloat[Filling factor $n = 1/4$]{\includegraphics[width=0.5\columnwidth]{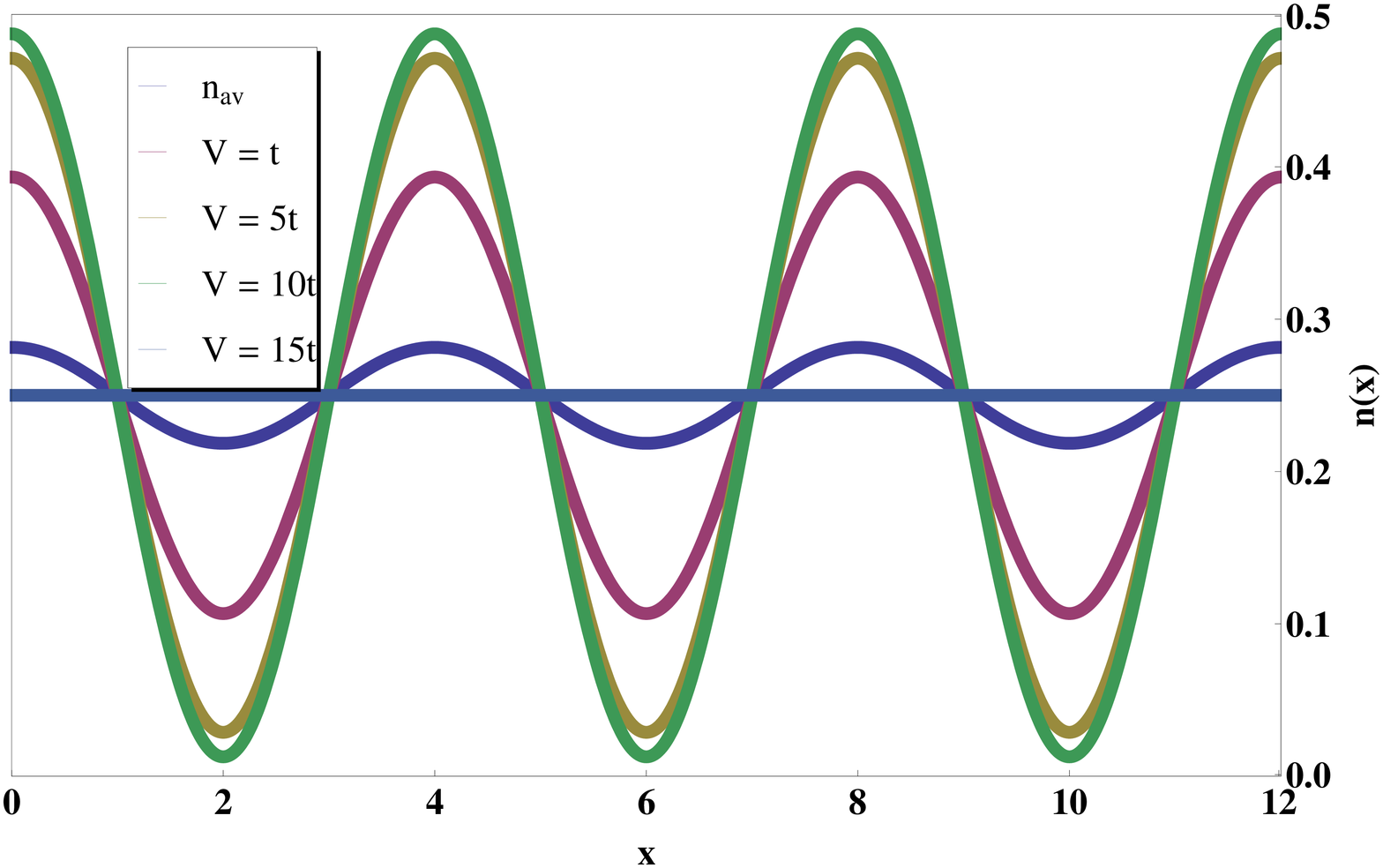}}
\caption{[Color online] The density variation in the Wigner crystal phase for two different filling factors.} \label{den}
\end{figure*}

Since the Bragg peak at $q = Q$ corresponding to the periodic ordering of the Wigner crystal phase, the average density distribution of the lattice can be written as $n(x) = n + I_s \cos (Q x)$. This quantity for different interaction parameters is shown in FIG. \ref{den}. As can be seen from the figure for both $n = 1/2$ and $n = 1/4$ filling factors, the higher interactions enhance the peak structure showing the crystalline structure in the density distribution. This periodic density order can be probed by using a currently available experimental technique, known as quantum gas microscopy \cite{nate, mark, imman}.

As we discussed in Sec. V, for a more realistic experimental setups, one has to consider the inter-chain interaction in the form given in Eq. (\ref{e10}). For $\theta_d =0$ intra-tube interaction is repulsive, however fermions in different tubes can attract or repel depending on their dipolar moment alignment and the tube separation. For attractive inter-tube interactions, the system forms a clustered Wigner crystal phase~\cite{n7}. This phase is coherent and Wigner crystals in both tubes locked to each other. On the other hand, for repulsive inter-tube interactions, the Wigner crystal phase is in in-coherent state.

As we mentioned before, our variational wave function always represents a Wigner crystal phase. In order to justify the phase transition between a liquid state and a Wigner crystal phase, here we compare the energy of the Wigner crystal phase and the liquid phase perturbatively. Taking the unperturbed wave function as a free particle state, the liquid state energy in the first order perturbation is given by

\begin{eqnarray}
E_l &=& -\frac{2t}{\pi}\sin(Q/2) + \frac{1}{2L}\biggr[V(0)\biggr(\frac{Q}{2\pi}\biggr)^2-\sum_q\frac{V(q)}{2\pi}(Q-q)\biggr].\label{lse}
\end{eqnarray}

\noindent By comparing the liquid state energy $E_l$ and the Wigner crystal state energy $E_g$, we find that the liquid state is favorable for small $V$ values and phase transition between these states takes place at a finite $V$ values for all filling factors. For example, for the case of quarter filling case, we find the phase transition at $V = 4.74 t$.

\section{VII Summary}

We have studied dipolar fermions in a one-dimensional lattice using the bosonization theory and a variational approach. In the weak coupling limit at half-filling, the bosonization theory predicts the appearance of several quantum phases as one change the polarization direction of the dipoles relative to the one-dimensional lattice orientation. The quantum phase diagram includes a charge-density wave, a spin-density wave, a singlet superfluid, and a triplet superfluid phases. In the strong coupling limit at lower filling factors, our variational method predicts the emergence of a Wigner crystal phase due to long-range interaction. The structure factor and the density distribution clearly indicates the existence of Wigner crystal at larger interactions. The entire rich phase diagram resulting from the competition between kinetic energy and the on-site and off-site long-range interactions can be detected by using currently available experimental techniques.

\section{VIII ACKNOWLEDGMENTS}

We thank Erik Weiler for carefully reading the manuscript.

\end{document}